\documentclass [manuscript]{aastex}

\usepackage{amssymb, amsmath, amsfonts, graphicx, multirow, epstopdf}

\newcommand{\vp}{\varphi}

\begin{document}

\title{Charging of Aggregate Grains in Astrophysical Environments}
\author{Qianyu Ma, Lorin S. Matthews, Victor Land \altaffilmark{1}, and Truell W. Hyde}
\affil{Center for Astrophysics,Sspace Physics, and Engneering Research, Baylor Univeristy, Waco, TX 76798}
\altaffiltext{1}{Foundation for Fundamental Research on Matter, Utrecht, The Netherlands}

\begin{abstract}
The charging of dust grains in astrophysical environments has been investigated with the assumption these grains are homogeneous spheres. However, there is evidence which suggests many grains in astrophysical environments are irregularly-shaped aggregates. Recent studies have shown that aggregates acquire higher charge-to-mass ratios due to their complex structures, which in turn may alter their subsequent dynamics and evolution. In this paper, the charging of aggregates is examined including secondary electron emission and photoemission in addition to primary plasma currents. The results show that the equilibrium charge on aggregates can differ markedly from spherical grains with the same mass, but that the charge can be estimated for a given environment based on structural characteristics of the grain.  The "small particle effect" due to secondary electron emission is also important for determining the charge of micron-sized aggregates consisting of nano-sized particles.
\end{abstract}

\keywords{dust, extinction --- interplanetary medium --- ISM:general --- solar system: general}

\section{Introduction}

Dust, an ubiquitous component in the universe, plays an important role in the thermodynamics and chemistry of the interstellar and intergalactic medium, interstellar gas dynamics, and the formation of stars, planets and planetesimals \citep{Jones97}. Cosmic dust grains also interact with electromagnetic radiation nearby, altering the observed spectra of remote objects \citep{Draine03}. The study of cosmic dust has steadily gained attention as technological advances makes \textit{in-situ} measurements and sample-collection within the heliosphere feasible. Various space missions have been conducted to probe and investigate the composition, size distribution and structural characteristics of interstellar and interplanetary dust, such as Ulysses, Cassini and Galileo \citep{Altobelli03,Krueger10a,Krueger10b}. These dust grains provide an excellent window into interstellar and interplanetary processes, carry information on the origin and evolution of their parent bodies, and reveal the intrinsic properties of the environments where these grains originate. The \textit{in-situ} measurements also provide opportunities to test and validate various theories related to cosmic dust.  

Dust grains in the solar system environment are subject to three charging processes \citep{Mendis94} --- for dust grains close to the Sun, strong UV radiation can excitep and liberate photoelectrons from the surface of the dust grain and charge the grain positively. At the same time, free moving electrons and ions in the solar wind constitute plasma charging currents, although the strength of the currents may be less than the photoelectric charging current \citep{Kimura98a}. Between the termination shock and the heliopause, the boundary of the heliosphere, the plasma temperature rises to $2 \times 10^6$ K as a result of the constant interaction between the interstellar medium and the solar wind. Secondary electron emission then becomes the dominant charging process \citep{Kimura98a}.

Ascertaining the charge on cosmic dust grains in the solar system is essential for several reasons. First, charged grains are subject to the Lorentz force, which can significantly alter their trajectories. When performing on-board measurements and sample-collection, the grain dynamics must be taken into account to identify the origin of the measured/captured grains. Second, the high plasma temperature between the termination shock and the heliopause highly charges interstellar dust grains entering the solar system. The smaller mass grains are affected most strongly by the Lorentz force caused by the magnetic field at the heliopause. As a result, their trajectories are significantly altered and these grains have much smaller probability of reaching the inner solar system. Space missions devoted to studying the interstellar dust in the inner solar system have to take this filtering effect into account \citep{Frisch99,Landgraf00,Linde00}. Third, grain charging theory has successfully explained many astrophysical phenomena, such as the mysterious spokes of Saturn's B ring, and the lunar ``horizon glow'' due to dust levitating above the moon's surface \citep{Mitchell06, Stubbs06}.

Charging of cosmic dust grains has been examined in detail due to the reasons listed above \citep{Feuerbacher73, Draine79, Mukai81,Chow93, Kimura98a}. However, most of the previous work assumes a simple geometry for dust grains, such as a homogeneous sphere. Dust grains naturally occurring in space constantly go through formation processes such as nucleation, condensation, coagulation and destruction, and as a result they assume more complex structures such as ellipsoids or fluffy aggregates. Aggregates are a common structure for interplanetary, cometary and interstellar dust. The data and samples collected from space have shown that a porous fluffy structure, consisting of many small subunits, can be found among interplanetary dust grains and in cometary bodies \citep{Brownlee80, Greenberg90, Hu11}. Although no direct evidence shows that interstellar dust grains assume the same structure, data analysis from remote observation supports the existence of aggregate structure among interstellar dust grains \citep{Jones88,Woo94,Wurm03}. The charging of aggregate dust grains has been studied recently both in laboratory and astrophysical environments \citep{Wiese10,Ilgner12,Okuzumi12}; these results have shown that aggregates tend to acquire more charge when compared to spherical grains of the same mass due to the porous/fluffy structure of the aggregate, and that charged aggregates have a significant effect on subsequent dust evolution. However,  there has not been a detailed study of characterizing the charge on aggregates based on structural characteristics. 

In this paper, a 3D model is employed to calculate the charge on aggregate dust grains under charging conditions particular to solar system environments. Three different charging processes are taken into account --- plasma charging currents, secondary electron emission and UV radiation. For charging processes dominated by secondary electron emission at the heliopause, we show a charge enhancement for all the aggregates compared to spheres of the same mass.  This charge-to-mass ratio is significantly higher for aggregates consisting of nano-sized grains as a result of the small particle effect. For photoemission, the charge on aggregates may be more or less than the charge on spheres with the same mass, depending on the magnitude of the variable solar UV photon flux.  Charge estimate models are proposed for both secondary electron emission and photoemission, and it is shown that the charge on aggregates can estimated based on structural characteristics such as the compactness factor.

\section{Charging Model}

The charge on a dustl grain embedded in plasma is determined by
\begin{equation}\label{eq:equi_charge}
\frac{dQ}{dt} = \sum_j I_j,
\end{equation}
where $I_j$ is the current contributed by the $j^{th}$ charging process. The charge on the grain reaches equilibrium when $\sum_j I_j = 0$. For an isolated spherical grain, Equation~\ref{eq:equi_charge} can be solved analytically to yield the equilibrium charge \citep{Goertz89}. However, this highly idealized circumstance is almost never satisfied in astrophysical environments. As discussed above, the cosmic dust grains often assume irregular shapes, thus require numerical simulation. The charging current density for three charging processes --- plasma charging, secondary electron emission and photoelectric emission, are given below. Silicates have been identified as one of the major constituents for cosmic dust grains \citep{Savage79,McCarthy80}, thus silicate grains with a density of 3.2 g $\mathrm{cm}^{-3}$ \citep{Draine79} are used as the grain material in this study to demonstrate the different charging processes.

\subsection{Collection of Plasma Particles}
The current density to a spherical grain can be found from Orbital Motion Limited theory (OML), which is based on the conservation of energy and angular momentum \citep{Whipple81, Laframboise73}. The current density to any point on the surface of a grain due to the collection of a given species of plasma particles is given by
\begin{equation}\label{eq:plasma_charge}
J_s = n_sq_s \int \!\!\! \int \!\!\! \int v_sf(\vec{v}_s)\cos\alpha d\vec{v}_s^3,
\end{equation}
where $n_s$ and $q_s$ are the number density and charge of the given species, $v_s$ is the speed of the particles, $f(\vec{v}_s)$ is the distribution function which is assumed to be Maxwellian \citep{Goertz89}, and $\alpha$ is the angle between the impinging velocity and the surface normal of the dust grain. In the three dimensional case, we use spherical coordinates $(v, \theta, \phi)$ in $\vec{v}$ space \citep{Laframboise73}. The differential velocity $d\vec{v}_s^3$ can be written as
\begin{equation}\label{eq:vec}
d\vec{v}_s^3 = v^2dvd\Omega,
\end{equation}
This allows the integration over the speed to be separated from the integral over the open solid angles, allowing Equation~\ref{eq:plasma_charge} to be rewritten as
\begin{equation}\label{eq:plasma_current}
J_s = n_sq_s \int_{v_{min}}^{\infty}v_s^3f(v_s)dv_s\int \!\!\! \int \cos \alpha d\Omega.
\end{equation}
The integration over speed is easy to carry out with $v_{min}$ given by
\begin{equation}\label{eq:vec_limit}
v_{min}= \begin{cases} 0, & q_s \varphi \geq 0 \\
\displaystyle \sqrt{\frac{-2q_s\varphi}{m_s}}, & q_s \varphi < 0\end{cases} ,
\end{equation}
where $\varphi$ is the surface potential of the grain and $m_s$ is the mass of the plasma particle. For a point on the surface of an isolated sphere, the integral over the open solid angles (a hemisphere) is also simply evaluated.  However, on the surface of an aggregate, not all of the incident angles are open to the incoming particle flux.  Thus, the differential solid angle $d\Omega$ requires numerical simulation for aggregates, which is discussed in Section 2.4.

\subsection{Secondary Electron Emission}
Energetic primary electrons can release secondary electrons from the surface of a grain upon impact, which constitutes a positive charging current. It has been shown that the secondary electron yield is enhanced when the dimensions of the monomers are comparable to the primary electron penetration depth, the so-called small particle effect \citep{Chow93}. Since the size of a representative interstellar dust grain is normally less than 10 $\mu$m, we employ a model which takes the small-particle effect into account in determining the yield, $\delta$, as a function of $E_0$, the initial energy of the primary electron \citep{Draine79},
\begin{equation}\label{eq:sec_eng}
\delta (E_0) = \delta_m \frac{8E_0/E_m}{(1+E_0/E_m)^2}{\left[1-exp{\left(\frac{-4a}{3\lambda}\right)}\right]}f_1{\left(\frac{4a}{3R}\right)}f_2{\left(\frac{a}{\lambda}\right)}.
\end{equation}
Here
\begin{align}
f_1(x) &= \frac{1.6+1.4x^2+0.54x^4}{1+0.54x^4} \nonumber \\
f_2(x) &= \frac{1+2x^2+x^4}{1+x^4},
\end{align}
and $a$ is the radius of the grain. The maximum yield $\delta_m$, and the corresponding maximum energy $E_m$, are 2.4 and 400 eV for silicates \citep{Mukai81}. The escape length $\lambda$ is 2.3 nm \citep{Draine79}. The projected range $R$ gives the penetration depth of a primary electron into matter along the incident direction, and is determined based on $E_0$ as shown by Draine and Salpeter \citep{Draine79}.

Thus, the current density due to secondary electron emission is calculated as
\begin{equation}\label{eq:sec_charge}
J_s = n_eq_e \int \!\!\! \int \!\!\! \int vf(v)\cos\alpha \delta(E_0) d\vec{v}^3 \times \int_{E_{min}}^{\infty} \rho(E)dE,
\end{equation}
where $\rho(E)$ is the energy distribution of the emitted electrons. It can be written as
\begin{equation}\label{eq:sec_rho}
\rho(E) = \frac{E}{2(kT_{sec})^2}{\left[1+\frac{1}{2}{\left(\frac{E}{kT_{sec}}\right)}^2\right]}^{-3/2},
\end{equation}
where $T_{sec}$ is the temperature of the emitted electrons and is set to be 2 eV \citep{Goertz89}. The lower limit of the integral is $E_{min} = max(0,e\varphi)$, with $\varphi$ being the surface potential of the target grain. Equation~\ref{eq:vec} can be used in \ref{eq:sec_charge} to yield
\begin{equation}\label{eq:sec_current}
J_{sec} = n_eq_e \int_{v_{min}}^{\infty} v^3f(v)\delta(E_0) dv \int \!\!\! \int \cos\alpha d\Omega \int_{E_{min}}^{\infty} \rho(E)dE,
\end{equation}
which has a form similar to that of Equation~\ref{eq:plasma_current}, with the only term dependent on the geometry of the aggregate being the integral over $d\Omega$. 

\subsection{Photoelectric Emission}
Incoming photons with energy $h\nu > W$, the work function of the material, can excitep and liberate electrons from the surface, and thus constitute a positive charging current. Assuming an isotropic source of UV, the photoelectric current density can also be separated into integration over the photon energy and the incident angles \citep{Kimura98a}:
\begin{equation}
J_{ph} = q_e \int_{W}^{\infty} Q_{abs}(h\nu) F(h\nu) Y(h\nu)\, d(h\nu)\,\int \!\!\!\int \cos\alpha d\Omega \int_{E_{min}}^{E_{max}} f(E) \, dE,
\end{equation}
where $h\nu$ is the photon energy and $F(h\nu)$ is the photon flux at a given distance from the Sun, which can be easily obtained through satellite measurements. The absorption efficiency $Q_{abs}$ depends on the grain radius $a$ and the wavelength of the incoming photon, $\lambda$. If $2\pi a > \lambda$, the interaction of the grain and the photon can be regarded as elastic scattering and $Q_{abs} = 1$. If $2\pi a < \lambda$, Mie scattering is often used to obtain $Q_{abs}$ \citep{Bohren83}. The work function of silicates, $W$, is 8 eV based on empirical value \citep{Draine79}. The photoelectric yield $Y(h\nu)$  is estimated using \citep{Draine79}:
\begin{equation}
Y(h\nu) = \frac{(h\nu-W+\epsilon_{min})^2-\epsilon_{min}^2}{(h\nu)^2- \epsilon_{min}^2}{\left\lbrack 1-(1-\frac{l_e}{a})^3 \right\rbrack},
\end{equation}
where the escape length of the photoelectrons $l_e$ is 1 nm for silicates \citep{Draine79}. $\epsilon_{min}$ is the minimum energy needed for photoelectric emission to occur, and is set to 6 eV for silicates \citep{Draine79}.

The energy distribution of the photoelectrons is given by $f(E)$ and must be taken into account if the potential distribution about the surface is such that some of the photoelectrons return instead of escaping. Both laboratory and space experiments indicate that the photoelectrons are emitted isotropically with a Maxwellian distribution for the energy at a characteristic temperature of 1-2 eV \citep{Hinteregger59,Grard73,Wrenn73}. Thus, the energy distribution $f(E)$ of the photoelectrons is:
\begin{equation}
f(E) = \frac{E}{(kT_{ph})^2}\exp{\left(-\frac{E}{kT_{ph}}\right)},
\end{equation}
with temperature $T_{ph}$ of the photoelectrons set to be 1 eV in the current study. The lower limit $E_{min}$ of the integration is $\max(0, e\vp_s)$. Thus the integration yields
\begin{equation}\label{eq:photo_e_rho}
\int_{E_{min}}^{\infty}f(E)\, dE = \begin{cases} 1, & q_e\varphi_s <0 \\
                                                 \displaystyle \exp{{\left(-\frac{q_e\varphi_s}{kT_{ph}}\right)}}, & q_e\varphi_s \geq 0.
                                    \end{cases}
\end{equation}

With the aid of Equation~\ref{eq:photo_e_rho} the photoelectric current density can be written as
\begin{equation}\label{eq:photo_current_dens}
\begin{cases} J_{ph} = q_e \displaystyle \int_W^{\infty}Q_{abs}(h\nu)F(h\nu)Y(h\nu) \, d(h\nu)\, \displaystyle \int \!\!\!\int \cos\alpha d\Omega, & q_e\varphi_s < 0\\
              J_{ph} = q_e exp{\displaystyle {\left(-\frac{q_e\varphi_s}{kT_{ph}}\right)}} \displaystyle \int_W^{\infty}Q_{abs}(h\nu)F(h\nu)Y(h\nu) \, d(h\nu) \,\displaystyle \int \!\!\!\int \cos\alpha d\Omega \, & q_e\varphi_s \geq 0.
\end{cases}
\end{equation}

Once again, the current density depends on the aggregate geometry through the integral over the open angles.  Since only the side of an aggregate currently facing the sun is illuminated, the photon current is estimated by dividing by a factor of two (assuming isotropic flux).  This should give an upper bound for the photoelectric current as the rotational period of a micron-sized dust grain is very short compared to the equilibrium charging time (see Section 3.2).

\subsection{Line-of-Sight Approximation}

\begin{figure}[h]
\begin{center}
\includegraphics[scale = 0.6]{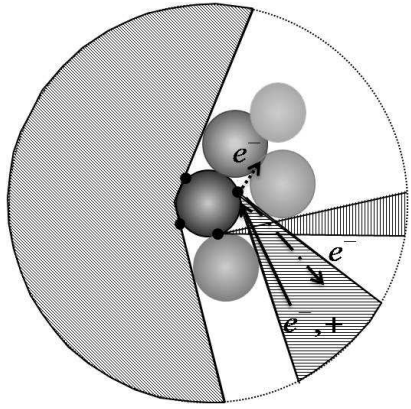}
\end{center}
\caption{Open lines of sight to given points on a monomer in an aggregate are indicated by the shaded regions. Charging currents to a given point are only incident from these directions. The dotted line indicates an emitted electron which is recaptured by another monomer along a closed line of sight, while the dash-dotted line indicates an emitted electron that escapes along a free line of sight.}
\label{fig:line_of_sight}
\end{figure}

The charging code OML\_LOS calculates the electron and ion fluxes by determining the open lines of sight (LOS) to the points on the surface of each constituent monomer. A detailed description can be found in \citet{Matthews12}; here a brief summary is given. Electrons and ions coming from the surrounding plasma are assumed to move in straight lines and are captured at the points at which their straight line trajectory intersects a monomer, as illustrated in Figure~\ref{fig:line_of_sight} .

The surface of each monomer is divided into many equal-area patches. Test directions $t$ from the center of each patch (the so called lines of sight) are determined to be \textit{blocked} if they intersect any other monomer in the aggregate, or the monomer in question ($\mathrm{LOS}_t = 0$), and \textit{open} otherwise ($\mathrm{LOS}_t = 1$). The integration over the angles in Equations~\ref{eq:plasma_current}, \ref{eq:sec_current} and \ref{eq:photo_current_dens} is replaced by the line-of-sight factor, which is equal to the sum of the open lines of sight multiplied by the cosine of the angle of the test direction with respect to the surface normal and by the area of the patch on a unit sphere, $\mathrm{LOS} = \int \!\!\!\int \cos \alpha d\Omega = \sum_t \mathrm{LOS}_t \cos \alpha_t \Delta(\cos\theta)\Delta\phi$. The net current of species $s$ to a given patch at a given time, $I_s(t)$, is found by multiplying the current density by the area of the patch, $A$: $I_s(t) = J_s(t)A$. Summing over species $s$ provides the change in the surface charge on the patch during a time interval $dt$, $dQ(t) = \sum I_s(t)dt$. The contribution to the dipole moment is given by $d\mathbf{D}(t) = \sum I_s(t)\mathbf{R}dt$, where $\mathbf{R}$ is the displacement vector from the patch to the center of the grain. The current density $J_s(t)$ depends on the potential at the center of the given patch, which in turn depends on the charge and dipole moment on each monomer,
\begin{equation}
V_{patch} = \frac{1}{4\pi\epsilon_0}{\left(\sum_i \frac{Q_i}{r_i} + \sum_i \frac{\displaystyle \mathbf{D}_i \cdot \displaystyle \mathbf{r}_i}{|\displaystyle \mathbf{r}_i|^3}\right)},
\end{equation}
where $Q_i$ and $\mathbf{D}_i$ are the charge and dipole on the \textit{i}th monomer, and $\mathbf{r}_i$ is the distance vector from the center of the $i^{th}$ monomer to the patch. The solution requires numerical iteration until equilibrium is reached. The change in the charge and dipole moment of each monomer is then obtained by adding the contribution of all the patches. The change in the charge and dipole moment of the aggregate is obtained by adding the contribution from each of the $N$ monomers. This process is iterated in time until the average change in aggregate charge becomes negligible, $dQ_{agg} < 0.0001Q_{agg}$, at which point the net current to the aggregate will be near zero.

In computing the current due to secondary electron emission or photoemission, an electron escapes from the aggregate only if the randomly chosen escape direction is along an open line of sight. Electrons which are released along a blocked line of sight are recaptured by another monomer within the aggregate, leaving the total charge of the aggregate unchanged, but the charge distribution on the surface is altered.

\subsection{Aggregate Builder and Compactness Factor}

The numerical code \textit{Aggregate Builder} was used to create aggregates through the coagulation of spheres using a combination of particle-cluster aggregation (PCA), and cluster-cluster aggregation (CCA) \citep{Matthews07,Matthews09}. During PCA, a target particle is placed at the origin, and a single particle is released at the boundary of the simulation box with its velocity directed towards the center of target particle plus an offset. A successful collision is detected if constituents of the target and projectile actually touch or overlap. The grains are assumed to have relative velocities that are too low for any restructuring to occur, and to stick at the point of contact \citep{Wurm98,Blum00}. New aggregate parameters are then calculated, and the resultant aggregate is saved to a library. In the case of CCA, small aggregates from the previously saved library are employed as the target grains, with the incoming grain either a spherical monomer or an aggregate randomly selected from the same library. 

While the structure of the aggregates (characterized by the compactness factor, described below) depends on the plasma environment in which it grows, the charge on an aggregate within a given environment is in turn a function of the compactness factor \citep{Matthews12}. Thus  a large number of aggregates were built covering a wide range of compactness factors assuming a neutral environment. The aggregates from the library were then charged through OML\_LOS using the parameters representing different astrophysical environments.

\begin{figure}[h]
\begin{center}
\includegraphics[scale = 0.6]{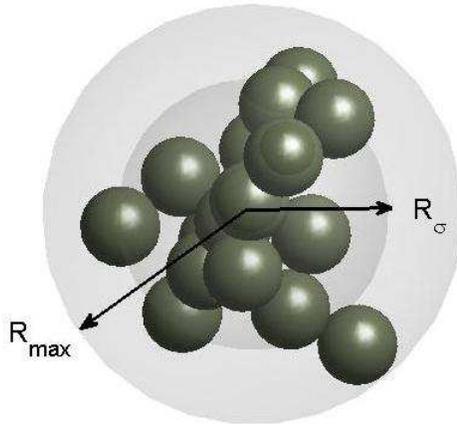}
\end{center}
\vspace{-10pt}
\caption[Illustration of the compactness factor]{Illustration of the compactness factor. The inner shaded area corresponds to a sphere with radius $R_{\sigma}$, the outer shaded area to a sphere with maximum radius, $R_{max}$. }
\label{fig:Compactness}
\end{figure}

The compactness factor, $\Phi_{\sigma}$, defined by Paszun and Dominik \citep{Paszun09}, is used to characterize the structure or fluffiness of an aggregate consisting of spherical monomers.
\begin{equation}
{\Phi}_{\sigma} = N{\left(\frac{a}{R_{\sigma}}\right)}^3,
\end{equation}
where $N$ is the number of monomers in the aggregate, $a$ is the constituent monomer radius, and $R_{\sigma}$ is the radius of the average projected surface area, defined as
\begin{equation}
R_{\sigma} = \sqrt{\frac{\sigma}{\pi}},
\end{equation}
with $\sigma$ being the projected surface area averaged over many orientations. Figure~\ref{fig:Compactness} shows a representative aggregate consisting of mono-disperse monomers, with $R_{\sigma}$ and the outer radius $R_{max}$ indicated. For compact aggregates, the volumes of the two spheres with these radii are approximately equal. For open aggregates, the ratio of the two approaches zero.

\begin{figure}[h]
\begin{center}
\includegraphics[scale = 0.3]{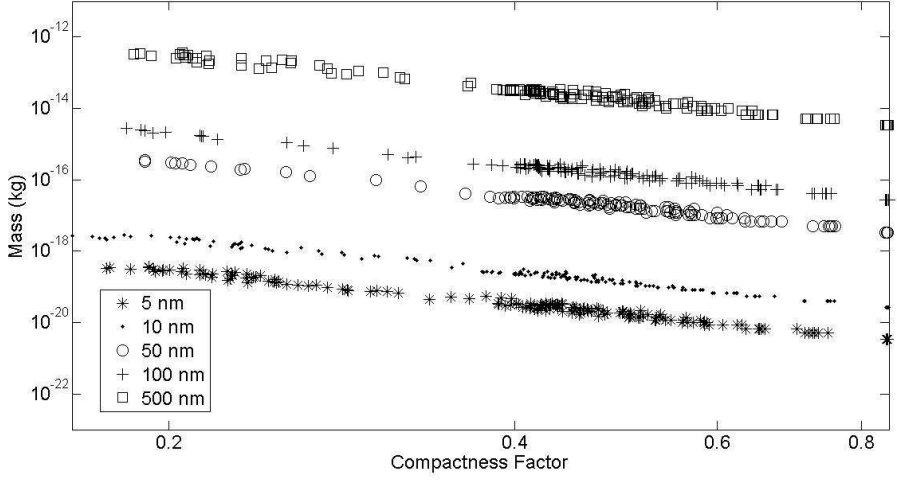}
\end{center}
\vspace{-10pt}
\caption{Mass of aggregates as a function of the compactness factor. The radius of the constituent monomers is given in the legend.}
\label{fig:Comp_mass}
\end{figure}

Figure~\ref{fig:Comp_mass} shows the log-log plot for the mass of aggregates consisting of monodisperse monomers of different radii (5 nm $\leq a \leq$ 500 nm), ranging in size from two to 200 monomers, as a function of the compactness factor. For each of the groups, as the mass increases, the compactness factor decreases, indicating a fluffier structure. This power-law relationship confirms that the compactness factor can be used to characterize the structure of the aggregates.

\section{Results}

\subsection{Charging of Aggregates with Secondary Electron Emission}

In this section, the charging of aggregates is examined by including both plasma currents and secondary electron emission. The aggregates are charged using parameters for conditions in the heliosheath, the region between the termination shock and the heliopause. The plasma temperature in this region rises to $2 \times 10^6$ K, a result of the constant interaction between the solar wind and the interstellar medium. Thus, secondary electron emission is the dominant charging process due to the high electron temperature \citep{Kimura98a}. The plasma density and temperature at 150 AU are $n_e = n_i = 2 \times 10^5$ $\mathrm{m}^{-3}$ and $T_e = T_i = 2 \times 10^6$ K \citep{Schwenn90, Pauls96}. It is shown that the collective charge on an aggregate consisting of nano-sized grains is appreciably enhanced due to the small particle effect on each subunit. Two models for approximating charge on aggregates are proposed and the charge-to-mass ratio of the aggregates is compared to that of the spheres with the same mass.

\subsubsection{Charging Time}
Before estimating the equilibrium surface charge on the aggregates, the time to reach the equilibrium condition needs to be considered, for depending on the plasma parameters and the dynamic processes being considered, the equilibrium condition is not always satisfied for grains of all sizes.

\begin{figure}[h]
\begin{center}
\includegraphics[scale=0.25]{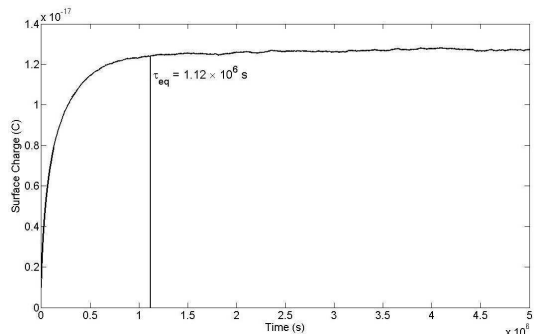}
\end{center}
\vspace{-10pt}
\caption{Charging curve for a dimer consisting of two 5 nm-radius monomers. $\tau_{eq}$ is determined by the point where the absolute change in the charge is less than 0.1\% of the equilibrium charge.}
\label{fig:dimer_charging_time}
\end{figure}

The dominant current determines the polarity of the equilibrium charge, while the non-dominant current determines the charging time, $\tau_{eq}$. As the grain charges, the relative contribution of the non-dominant current increases to balance the dominant current. Due to the high temperature of the plasma near the heliopause, secondary electron emission is the dominant charging process, determining $Q$. Thus $\tau_{eq}$ can be approximated by $|Q / I_e|$. Generally, $\tau_{eq}$ increases with decreasing dust radii $a$, approximately according to $\tau_{eq} \propto a^{-1}$. Figure~\ref{fig:dimer_charging_time} shows the charging history of a dimer consisting of two 5 nm-radius monomers, the smallest aggregate in the simulation. The maximum charging time is approximately $1 \times 10^6$ s, which is less than $5.75 \times 10^6$ s, the time needed for interstellar dust grains to travel 1 AU with a constant speed of 26 km/s. Since the typical distance between the heliopause and termination shock is 50 AU \citep{Schwenn90}, all the aggregates in the simulation are assumed to reach equilibrium within traveling a distance of 1 AU.

\subsubsection{Model for Estimating Aggregate Charge}

\begin{figure}[h]
\begin{center}
\includegraphics[scale=0.3]{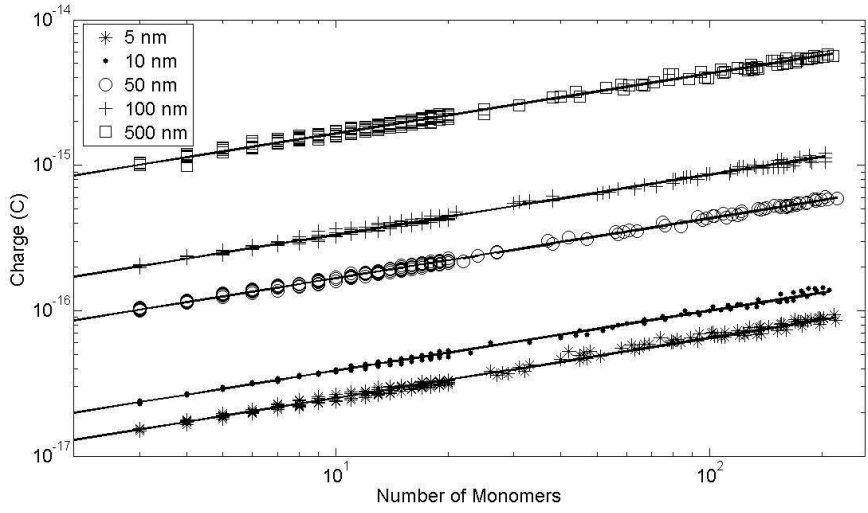}
\end{center}
\vspace{-10pt}
\caption{Surface charge on aggregates as a function of the number of constituent monomers. The linear fits have the same slope for all monomer sizes with standard error less than 2\%. The radius of the constituent monomers is indicated in the legend. }
\label{fig:sec_charge_num}
\end{figure}

\begin{figure}[h]
\begin{center}
\includegraphics[scale=0.3]{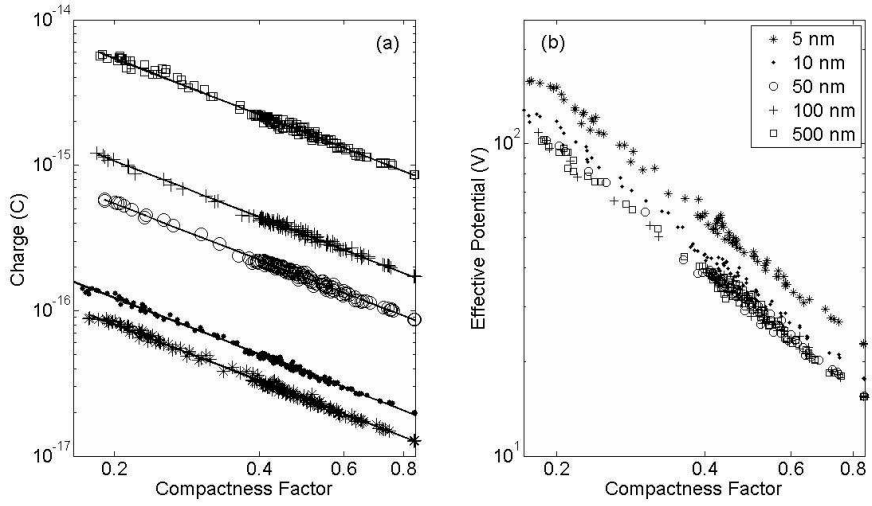}
\end{center}
\vspace{-10pt}
\caption{ (a) The charge on aggregates consisting of mono-disperse monomers of different radii, as indicated by the legend, and (b) the surface charge divided by the capacitance of a single monomer. The small particle effect is clearly evident for aggregates composed of monomers of size $a = 5$ nm.  }
\label{fig:sec_charge_comp}
\end{figure}

The equilibrium surface charge on aggregates is plotted using both the number of monomers (Figure~\ref{fig:sec_charge_num} ) and the compactness factor (Figure~\ref{fig:sec_charge_comp}a). The aggregates in each group consist of up to 200 mono-disperse monomers with radii ranging from 5 nm to 500 nm. Figure 5 shows that the aggregate charge is related to the number of monomers by
\begin{equation}\label{eq:q_N_sec_emission}
Q_{agg} \propto N^{0.413},
\end{equation}
where  $N$ is the number of monomers within an aggregate.

The aggregate charge can be predicted based on both the number of the monomers or the compactness factor. However, it is difficult, if not impossible, to determine the number of monomers within an aggregate measured \textit{in situ}, while compactness factor can be obtained through remote observation. In fact, much research has been devoted to relating the morphological structure of aggregates to their optical properties \citep{Kimura98b, Kimura01, Shen08}. As such, relating the charge to the compactness factor may serve as a useful tool when investigating the dynamics of interstellar dust grains in the outer heliosphere.

The equilibrium charge as a function of the compactness factor is shown in Figure~\ref{fig:sec_charge_comp}(a) for aggregates consisting of monodisperse spheres with different radii. Each group can be fit with a straight line of the same slope on a log-log plot with charge related to the number of constituent monomers by
\begin{equation}\label{eq:q_compfac_sec_emission}
Q_{agg} \propto \Phi_{\sigma}^{-1.3}.
\end{equation}
The results clearly demonstrate that for each group, the surface charge on the aggregate increases as the fluffiness of the aggregate structure increases, with the surface charge of a sphere being the lower limit as $\Phi_{\sigma}$ approaches one. This indicates that using the surface potential of a sphere of an equivalent mass to calculate the charge on an aggregate leads to charge underestimation.

In Figure~\ref{fig:sec_charge_comp}(b), the charges on the aggregates are divided by $4\pi\epsilon_0a$, the capacitance of a single monomer, to yield the effective potential for the aggregates. After eliminating the monomer size factor, it can be seen that aggregates consisting of monomers with $a > 10$ nm fall on the same line, while the aggregates with the smallest monomers, $a = 5$ nm, exhibit a substantially higher y-intercept. The aggregates consisting of monomers with $a = 10$ nm lie in between. It is evident that the collective contribution of the higher potential achieved by nm-radius grains (within the much larger aggregate) caused by the small particle effect is significant, and needs to be taken into account when estimating the charge on aggregate structures.

\begin{figure}[h]
\begin{center}
\includegraphics[scale=0.3]{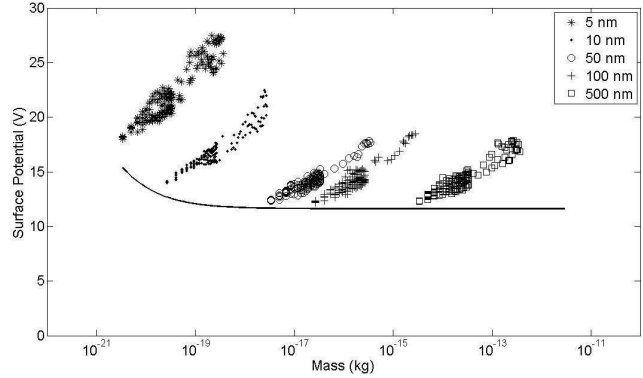}
\end{center}
\vspace{-10pt}
\caption{Comparison of the surface potential on aggregates (data points) and spheres having the same mass (solid line).  The monomer radius within the aggregates is indicated by the legend. }
\label{fig:surface_potential_vs_mass_helio}
\end{figure}

In a given plasma environment, spherical grains with $a > 10$ nm reach the same equilibrium potential, independent of their radii, while spheres with radii $a \leq$ 10 nm have a greater potential due to the small particle effect from secondary electron emissions \citep{Chow93}. However, as shown in Figure~\ref{fig:surface_potential_vs_mass_helio} , the surface potential of an aggregate clearly does not follow this trend. The surface potential of an aggregate in this case is defined as
\begin{equation}
\varphi = \frac{Q_{agg}}{ 4\pi\epsilon_0r_{mass}},
\end{equation}
where $Q_{agg}$ is the total charge on the aggregate and $r_{mass}$ is the radius of a solid silicate sphere having the same mass as the aggregate.  Overall, the surface potential of aggregates shows greater fluctuation and is generally greater than that of a sphere with the same mass, due to the greater surface area of the aggregate. A similar effect has been seen in an experimental study \citep{Wiese10}. Aggregates consisting of monomers with \textit{a} = 5 nm and 10 nm have a surface potential which is significantly higher than that for a sphere with the same mass. This is caused by the high positive charge each constituent monomer carries as a result of the small particle effect, consistent with prediction \citep{Kimura98a}.

\subsection{Charging of Aggregates with UV Radiation}

The typical plasma environment at 1 AU is used is used to illustrate charging by photoemission, as photoemission  is the dominant charging process at this distance \citep{Kimura98a}. Plasma and UV radiation parameters vary greatly over spatial distance and with time. However, the current purpose is to demonstrate the charging of aggregate grains compared to spherical grains, so more emphasis is placed on the characteristics of aggregate charging rather than modeling a specific environment. The plasma density and temperature at 1 AU are $n_e = n_i = 6 \times 10^6$ $\mathrm{m}^{-3}$ and $T_e = T_i = 2 \times 10^5$ K \citep{Schwenn90}. Only electrons and singly ionized hydrogen are considered, with other plasma components neglected due to their relatively small contribution \citep{Schwenn90}. Instead of simulating the photon flux in specific regions and time periods, the product of the yield and solar flux integrated over the spectrum is left as a free parameter and varied from  $5 \times 10^{12}$ $\mathrm{m}^{-2}\mathrm{s}^{-1}$ to $1.5 \times 10^{13}$ $\mathrm{m}^{-2}\mathrm{s}^{-1}$, within the photoemission current densities expected at 1 AU \citep{Whipple81}. The constituent monomers are taken to be silicate grains with radii $a = 50$ nm and 1 $\mu$m. The absorption efficiency $Q_{abs}$ is set to be unity, as the grain radius greatly exceeds the photon wavelength. The small particle effect for photoemission is also neglected due to the large radius of the grains.

\begin{figure}[!h]
\begin{center}
\includegraphics[scale=.8]{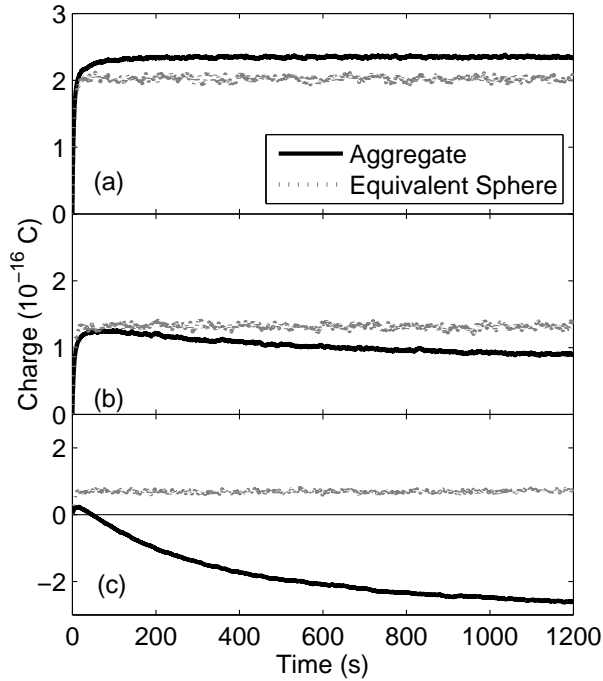}
\end{center}
\vspace{-10pt}
\caption{Evolution of charge on  aggregates compared to that of an equivalent sphere. The photoemission current density is $1.2 \times 10^{13}$ $\mathrm{m}^{-2}\mathrm{s}^{-1}$ in (a), $9 \times 10^{12}$ $\mathrm{m}^{-2}\mathrm{s}^{-1}$ in (b) and $6 \times 10^{12}$ $\mathrm{m}^{-2}\mathrm{s}^{-1}$ in (c).  The aggregate charge shown is the average for six different aggregates with $N$ = 13 monomers.}
\label{fig:UV_charge_on_aggs}
\end{figure}

In Figure~\ref{fig:UV_charge_on_aggs}, the time evolution of the aggregate surface charge is compared to that of an equivalent sphere for three different photoemission current densities of $6 \times 10^{12}$ $\mathrm{m}^{-2}\mathrm{s}^{-1}$, $9 \times 10^{13}$ $\mathrm{m}^{-2}\mathrm{s}^{-1}$ and $1.2 \times 10^{13}$ $\mathrm{m}^{-2}\mathrm{s}^{-1}$, respectively. The results indicate that aggregates and spheres may have charges of opposite polarity under the same conditions (Figure~\ref{fig:UV_charge_on_aggs}(c)), and may be either more or less highly charged than an equivalent sphere depending on the magnitude of the photoemission current density compared to the plasma current density, (Figure~\ref{fig:UV_charge_on_aggs}(a) and (b)). This is due to the porous structure of the aggregate. A highly irregular object has a greater surface area and is thus able to absorb more of the emitted electrons, as also shown in a recent experimental study \citep{Wiese10}. By the same token, when the photoemission current is very strong, the porous aggregate has more surface exposed to the UV photons, yielding a greater positive charge.  

\subsubsection{Model for Estimating Aggregate Charge}

\begin{figure}[!h]
\begin{center}
\includegraphics[scale=.8]{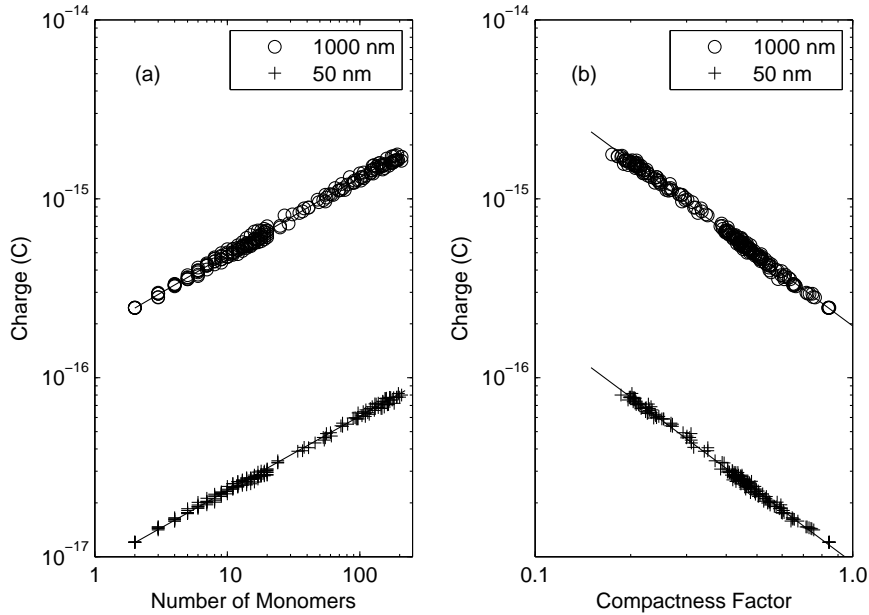}
\end{center}
\vspace{-10pt}
\caption{Surface charge on aggregates as a function of (a) the number of monomers and (b) the compactness factor. The radius of the constituent monomers is indicated in the legend.  The linear fits have the same slope for two sizes with standard error less than 3\%.}
\label{fig:charge_vs_N_UV}
\end{figure}

Again, we characterize the equilibrium charge on the aggregates due to plasma and photoelectric charging using both the number of monomers and the compactness factor. Using a photoemission current density of $1.1 \times 10^{13}$ $\mathrm{m}^{-2}\mathrm{s}^{-1}$, the charge on aggregates is plotted as a function of the number of monomers and the compactness factor in Figure~\ref{fig:charge_vs_N_UV}. Based on Figure~\ref{fig:charge_vs_N_UV} (a), the charge can be estimated as a function of the number of monomers,
\begin{equation}\label{eq:UV_charge_num}
Q_{agg} \propto N^{0.42},
\end{equation}
where $N$ is the number of the monomers within an aggregate. Figure~\ref{fig:charge_vs_N_UV} (b) indicates that a linear relationship on a log-log scale can also be obtained for the charge and the compactness factor,
\begin{equation}\label{eq:UV_charge_comp}
Q_{agg} \propto \Phi_{\sigma}^{-1.3}.
\end{equation}

The same exponential factor for both monomer sizes serves as strong evidence that aggregate charge is a function of the aggregate structure. The charge on aggregates can be estimated either by the number of the constituents, or the structural characteristics (fluffiness of the aggregate). While determining the charge on an aggregate based on the number of constituent monomers seems intuitive, the information is often hard or infeasible to obtain. Structure characteristics, on the other hand, can be obtained through the scattering and absorbtion interaction between aggregates and light. The power-law relation between the compactness factor and the aggregate charge also provides an indirect but a rather accurate method of determining the morphology of interplanetary dust. The Cosmic Dust Analyser (CDA) on the Cassini spacecraft has successfully detected the charge on interplanetary dust \citep{Kempf04}. If the composition and size distribution of these grains is known, along with the solar wind conditions, the structure of these grains may be obtained based on the charge estimate models proposed above.

\section{Discussion and Conclusion}

A numerical model has been used to calculate the charge on aggregate structures in astrophysical environments, including primary plasma currents, secondary electron emission, and photoemission.  It is shown that the charge on aggregates is strongly correlated to structural characteristics (Figure~\ref{fig:sec_charge_comp} and Figure~\ref{fig:charge_vs_N_UV}b, as measured by the compactness factor).  In general, porous aggregates, with their greater surface area, are more highly charged than an equivalent mass sphere, with the sign of the charge being determined by the dominant charging current.  The substantial increase in charge-to-mass ratio for aggregates in the region of the heliosphere (Figure~\ref{fig:surface_potential_vs_mass_helio}) will have a significant effect on the dynamics of these grains, greatly influencing the mass distribution of interstellar dust grains detected within the solar system by instruments such as the dust detector on Ulysses.  

It is interesting to note that the relationship between charge and structural factors (Equations~\ref{eq:q_N_sec_emission}, ~\ref{eq:q_compfac_sec_emission}, ~\ref{eq:UV_charge_num}, ~\ref{eq:UV_charge_comp}), is the same for the two different charging environments.  This is a result of the LOS factor for an aggregate being independent of the monomer size within the aggregate, as long as all of the spherical monomers have the same radius.  The relationship between charge and aggregate structural characteristics for polydisperse monomer populations is the subject of current research.  

Finally, the relationship between the charge on an aggregate and the charge on an equivalent sphere can vary greatly depending on the magnitude of the non-plasma currents.  This is seen for the aggregates charged by photoemission in Figure~\ref{fig:UV_charge_on_aggs}.  The values used for the photoemission current density in the three cases shown are all within the range expected for solar UV flux at 1AU, which varies greatly depending on solar activity.  Thus the charging history of aggregates in space can vary greatly over time, and is markedly different from the charging history of a spherical grain.  Further results exploring these differences will be presented in an upcoming paper.

\section{Acknowledgements}
This work was supported by NSF grant 0847127.

\bibliographystyle{plainnat}
\bibliography{disref}

\end{document}